# Two dimensional silicon chalcogenides with high carrier mobility for photocatalytic water splitting

Yun-Lai Zhu,[1+] Jun-Hui Yuan,[1+] Ya-Qian Song,[1] Sheng Wang,[1] Kan-Hao Xue,[1,2*] Ming Xu,[1] Xiao-Min Cheng,[1*] Xiang-Shui Miao[1]

[1]Wuhan National Research Center for Optoelectronics, School of Optical and Electronic Information, Huazhong University of Science and Technology, Wuhan 430074, China

[2]IMEP-LAHC, Grenoble INP – Minatec, 3 Parvis Louis Néel, 38016 Grenoble Cedex 1, France

[+]The authors Y.-L. Zhu and J.-H. Yuan contributed equally to this work.

## Corresponding Authors

*E-mail: xmcheng@hust.edu.cn (X.-M. Cheng)  xkh@hust.edu.cn (K.-H. Xue)

**Abstract:** Highly-efficient water splitting based on solar energy is one of the most attractive research focuses in the energy field. Searching for more candidate photocatalysts that can work under visible-light irradiation are highly demanded. Herein, using first principle calculations based on density functional theory (DFT), we predict that the two dimensional silicon chalcogenides, *i.e.* SiX (X=S, Se, Te) monolayers, as semiconductors with 2.43 eV~3.00 eV band gaps, exhibit favorable band edge positions for photocatalytic water splitting. The optical adsorption spectra demonstrate that the SiX monolayers have pronounced optical absorption in the visible light region. Moreover, the band gaps and band edge positions of silicon chalcogenides monolayers can be tuned by applying biaxial strain or increasing the number of layers, in order to better fit the redox potentials of water. The combined novel electronic, high carrier mobility, and optical properties render the two dimensional SiX a promising photocatalyst for water splitting.

## Introduction

It is imperative to develop technologies that can efficiently convert solar energy into renewable and clean energy sources, owing to the growing threat of energy crisis and environmental issues. Splitting water with solar energy to obtain hydrogen gas has been regarded as a promising candidate to this purpose.[1–9] The water splitting process involves a semiconductor photocatalyst that absorbs solar light, generating electron-hole pairs, whereby after charge separation $H_2$ can be generated with the aid of the electrons. However, there are several restrictions on the semiconductor, such as an appropriate band gap and favorable band alignments, in order to serve as an efficient visible-light photocatalyst.

Recently, a variety of two dimensional (2D) materials have been extensively studied as the photocatalysts for water splitting, due to their large specific surface areas as well as the short charge migration distances, which could enhance the catalytic performance by hindering the electron-hole recombination.[10–15] A typical example is monolayer $SnS_2$, which yielded a photocurrent density of 2.75 mA $cm^{-2}$ at 1.0 V, nearly 72 times larger than that of bulk $SnS_2$, proven in theory and experiment.[9] Other 2D materials such as transition metal dichalcogenides,[16] MXenes,[17] group-III monochalcogenides,[18] ternary zinc nitrides,[19,20] and $MPSe_3$[21] *etc.* have also been predicated theoretically for photocatalyst application.

Moreover, the booming research advancements of the stabilities and electronic properties of group IV–VI monolayers, which are isoelectronic counterparts of group V such as phosphorene, have been reported in the last few years.[22–32] The group IV mono-chalcogenides MX (M= Si, Ge, Sn and X = S or Se), whose buckled honeycomb lattice is similar to that of black phosphorene, are also candidate materials for photocatalytic water splitting. Nevertheless, their calculated overpotentials for OER are quite large, or a specific basic or acidic condition is required to obtain good photocatalytic activity.[33] The monolayer germanium monochalcogenides, like blue phosphorene, was predicted as UV-light-driven photocatalyst, owing to the large band gap.[34] Therefore, it is highly worthwhile to further investigate the electronic and optical properties of other group IV–VI monolayers, for the sake of finding new candidate materials with improved properties for optoelectronic devices.

Motivated by this conception, we have conducted a comprehensive investigation of the stability and electronic properties of silicon chalcogenides, *i.e.* SiX (X=S, Se and Te) monolayers, based on density functional theory. It is found that the SiX monolayers are of high dynamic, mechanical and thermal stability, favoring their experimental synthesis. Remarkably, the semiconducting SiX monolayers present a much larger strain-tunable band gaps, and possess considerable carrier mobilities. Furthermore, the band gaps and band edge positions of the SiX monolayers (bilayers) are suitable for photocatalystic water splitting, indicating their application potential.

## Computational methods

All density functional theory (DFT) calculations were carried out using the Vienna *Ab initio* Simulation Package (VASP). [35,36] The electron-ion interactions were described using the projector augmented-wave (PAW) method.[37,38] The generalized gradient approximation (GGA) within Perdew–Burke–Ernzerhof (PBE)[39] functional form was used for the exchange-correlation energy. The kinetic energy cutoff of the plane-wave basis was fixed to 500 eV. For all self-consistent calculations, the convergence criterion for total energy was set to $10^{-7}$ eV. In structural optimization, we relaxed the cells until the Hellmann-Feynman force acting on any atom was less than 0.005 eV/Å in each direction. The 2D Brillouin zone was sampled with a 15×15×1 Γ-centered *k*-point grid for geometric optimizations computations. A large vacuum space of ~20 Å in the perpendicular direction was introduced to get rid of the artificial interaction between layers and their periodic images. The effect of dipole correction was also included in our calculations. Since DFT-GGA usually underestimates the band gaps, the hybrid Heyd–Scuseria–Ernzerhof (HSE06)[35] functional was used to characterize the electronic band structures with higher accuracy. The DFT-D2 correction of Grimme[40] was adopted to describe the interlayer van der Waals (vdW) interaction for investigating the bilayer SiX. The phonon dispersion relations were calculated with the density functional perturbation theory as implemented in the PHONOPY code.[41] In addition, *ab initio* molecular dynamics (AIMD) simulations were performed to assess the thermodynamic stability of the structures, where NVT canonical ensembles were used.[42]

**Results and discussions**

The structure of SiX (X=S, Se, Te) monolayers can be regard as an analogue of silicene with Si atoms replaced alternately by chalcogens, as shown in **Fig. 1(a)**. The optimized lattice constants (3.30 Å for SiS, 3.52 Å for SiSe, 3.83 Å for SiTe), bond lengths (2.32 Å for SiS, 2.48 Å for SiSe, 2.69 Å for SiTe) and buckling heights (1.33 Å for SiS, 1.42 Å for SiSe, 1.53 Å for SiTe) of SiX monolayers demonstrate monotonic increase, following the same trend of the X atomic radii, as summarized in **Table 1**. This trend in buckling height is nevertheless different from that of GeS and GeSe configuration.[24]

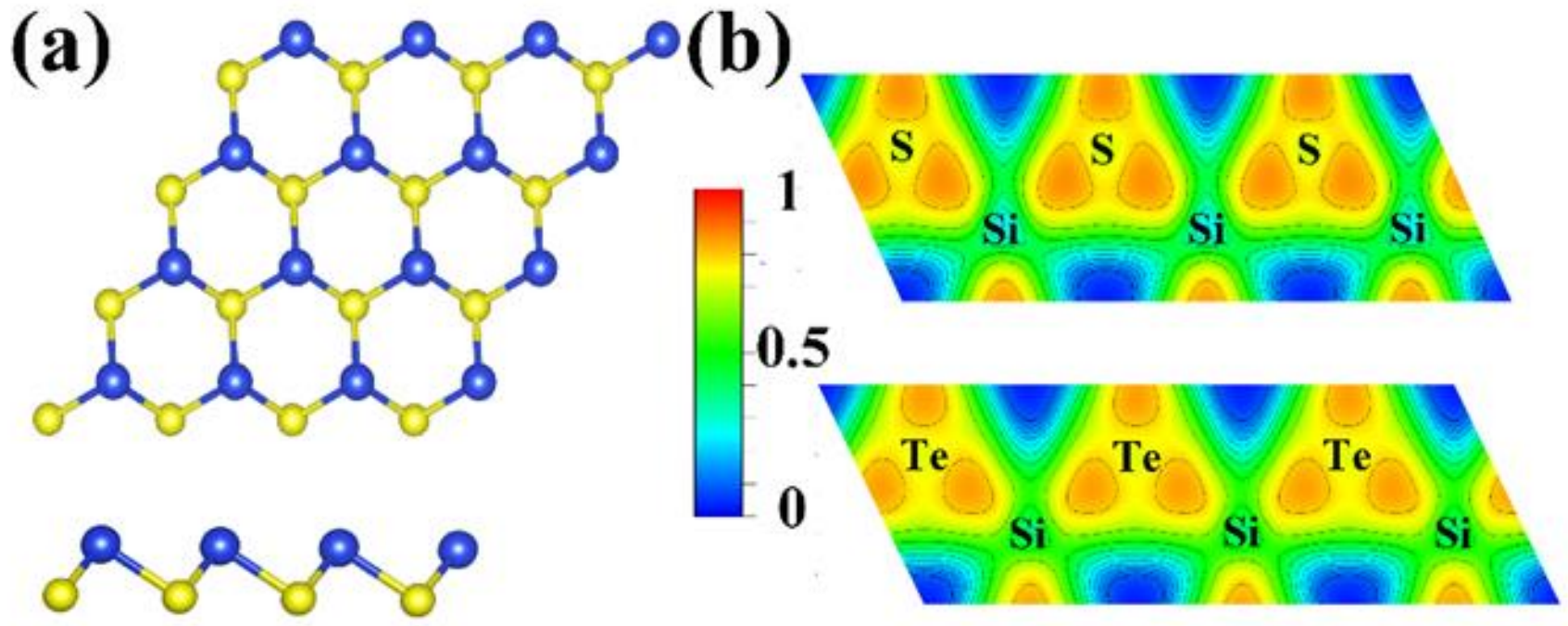

**Fig. 1** (a) Structure of monolayer SiX (X = S, Se, Te) from the top view and side view. The yellow and blue balls represent X and Si atoms, respectively. (b) Visualization of the electron localization function for monolayer SiS and SiTe.

**Table. 1** Calculated lattice constant $a$, buckling height $h$, cohesive energy $E_c$, band gaps $E_g$ (at both PBE and HSE06 levels) and charge transfer $T_e$ from the Si atom to the X atom of SiX monolayers.

| | $a$ (Å) | $h$ (Å) | $d_{Si\text{-}X}$ (Å) | $E_C$ (eV/atom) | $E_g$ (eV) PBE | $E_g$ (eV) HSE06 | $T_e$ (e) |
|---|---|---|---|---|---|---|---|
| SiS | 3.30 | 1.33 | 2.32 | 3.95 | 2.20 | 3.00 | 2.4 |
| SiSe | 3.52 | 1.42 | 2.48 | 3.60 | 2.12 | 2.90 | 2.3 |
| SiTe | 3.83 | 1.53 | 2.69 | 3.25 | 1.83 | 2.43 | 0.5 |

To investigate the chemical bonding features of SiX monolayers, the electron localization function (ELF) calculation[43–46] and Bader charge analysis [43,47,48] were performed. The ELF value ranges from 0 to 1, where ELF=1 indicates completely localization, ELF=0.5 corresponds to the case of free electron gas and ELF=0 represents the absence of electrons. As shown in **Fig.1 (b)**, some electrons have been transferred from Si to X, resulting in regions with high ELF values on the X atoms. Moreover, between Si and X atoms there are certain amount of electrons distributing continuously, with higher ELF values appearing near the X atom side, indicating the coexistence of ionic and covalent bonds for Si-X. The Bader charge analysis confirms the significant charge transfer from Si to X. As listed in Table 1, there are 2.4 e (2.3 e) charge transferred from Si to S(Se). Yet, the amount of charge transferred from Si to Te is only 0.5 e, which is consistent with previous results.[49] The large difference in charge transfer between SiS/SiSe and SiTe is mainly due to the discrepancy in electronegativity, *i.e.* 1.90 for Si, 2.58/2.55 for S/Se and 2.10 for Te, respectively. In addition, the cohesive energies $E_c$ of SiX are calculated using the expression $E_C = (nE_{Si} + nE_X - E_{SiX}) / 2n$, where $E_{SiX}$ is the total energy of the SiX monolayer, $E_{Si}$ and $E_X$ are the energies of the isolated Si and X atoms, and $n$ is the number of atoms. The calculated cohesive energies are 3.95 eV/atom, 3.60 eV/atom, and 3.25 eV/atom for SiS, SiSe and SiTe monolayers, respectively. These values are comparable to blue phosphorene (3.47 eV/atom) and higher than that of arsenene (2.96 eV/atom), demonstrating relatively strong binding in SiX.

For newly proposed 2D materials, stability is a critical aspect for experimental preparation and large-scale production. Thus, we have calculated the phonon dispersions to verify the kinetic stability of the SiX monolayers. No imaginary frequencies were observed, and linear dispersion relations around the Γ point were revealed, both indicating the structural rigidity and stability of SiX monolayers (**Fig. S1(a)-(c)**). In addition, the thermal stability of SiX monolayers was also examined by AIMD simulations for the NVT ensemble. A relatively large 5×5 supercell was used at room temperature (300K) for each material. The total simulation time was 5 *ps*, with a time step of 1 *fs*. The total energies of SiX monolayers oscillated with an extent of less than 0.04 eV/atom, and no obvious structural distortions were found (see **Fig. S1(d)-(f)**), suggesting that the structures of SiX monolayers are thermally stable at room temperature.

To further verify the mechanical stability of SiX monolayer, we also calculated the linear elastic constants, as listed in **Table S1**. The elastic constants of SiTe monolayer were calculated to be $C_{11}$ = $C_{22}$ = 35.53 N $m^{-1}$, $C_{12}$= 6.44 N $m^{-1}$, and $C_{44}$= 34.37 N $m^{-1}$, in good agreement with previous computations.[42] The stability criteria for a 2D hexagonal structure are[42, 43] $C_{11} > 0, C_{44} > 0, C_{11} - C_{12} > 0$, thus the mechanical stability of the SiX monolayers has been verified. Moreover, the in-plane Young's modulus ($Y$) and Poisson's ratio ($\upsilon$) for the SiX monolayers can be derived from the elastic constants by $Y = (C_{11}^2 - C_{12}^2)/C_{11}$ and $\upsilon = C_{12}/C_{11}$ , respectively.[50–52] The Young's moduli of SiX (X = S, Se, Te) monolayers are 51.27 N $m^{-1}$, 41.49 N $m^{-1}$, and 34.37 N $m^{-1}$, respectively, smaller than that of $MoS_2$ monolayer (129 $Nm^{-1}$)[53] and silicene (62 N $m^{-1}$).[54] Hence, the SiX monolayers are rather flexible materials, such that they be useful for practical large magnitude in-plane strain engineering.

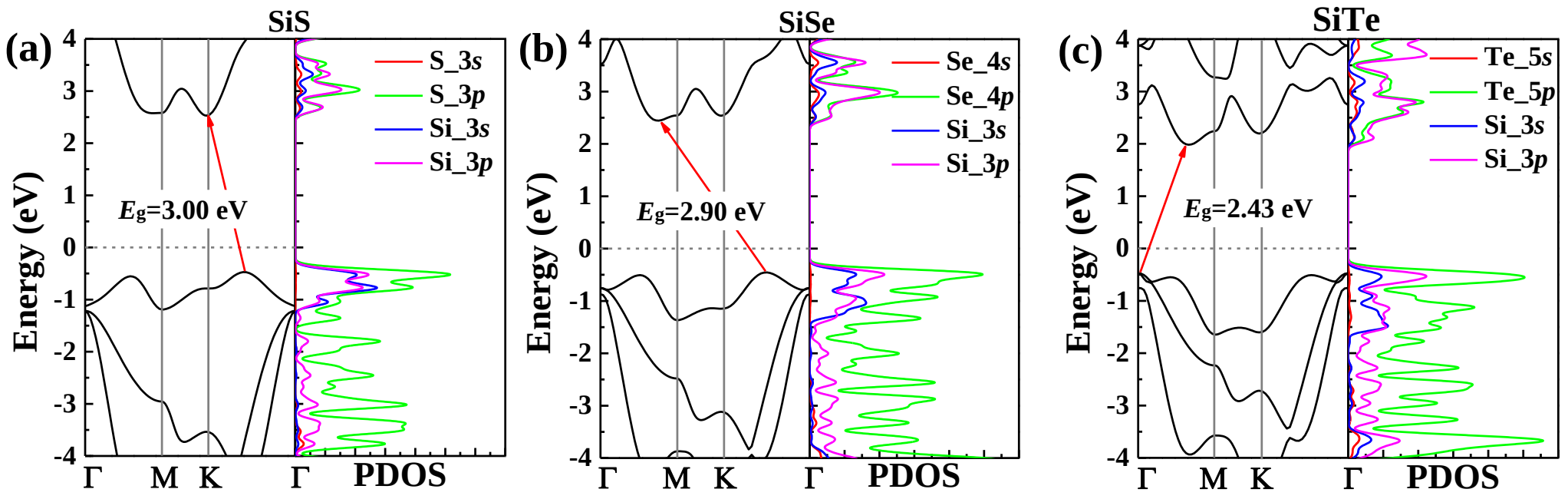

**Fig. 2** Calculated band structures and corresponding partial density of states (PDOS) of (a) SiS, (b) SiSe, and (c) SiTe monolayers, respectively, based on the HSE06 functional. The Fermi energy is set to zero.

After verifying the stability of SiX monolayers, we turn to their electronic structures. The SiX monolayers possess an indirect band gap feature, according to our PBE and HSE06 calculation results as plotted in **Fig. S2** and **Fig. 2**, respectively. The band gap values at the PBE level are 2.20 eV, 2.21 eV and 1.83 eV for SiS, SiSe and SiTe, respectively. Similar band structures but with larger indirect band gaps of 3.00 eV (for SiS), 2.90 eV (for SiSe) and 2.43 eV (for SiTe) have been confirmed based on the high accuracy HSE06 calculations. On the other hand, among these SiX monolayers one can find different energy band characteristics in terms of the locations of the conduction band minimum (CBM) and valence band maximum (VBM). For instance, the VBMs of SiS and SiSe reside nearly in the middle of $K$ and $\Gamma$ points, while it is exactly at the $\Gamma$ point for SiTe. Partial density of states (PDOS) analysis reveals that the VBMs of the SiX monolayers primarily consist of X-$p$, Si-3$s$ and Si-3$p$ states, while the CBMs mainly originate from the hybridization of Si-3*p* and X-*p* states.

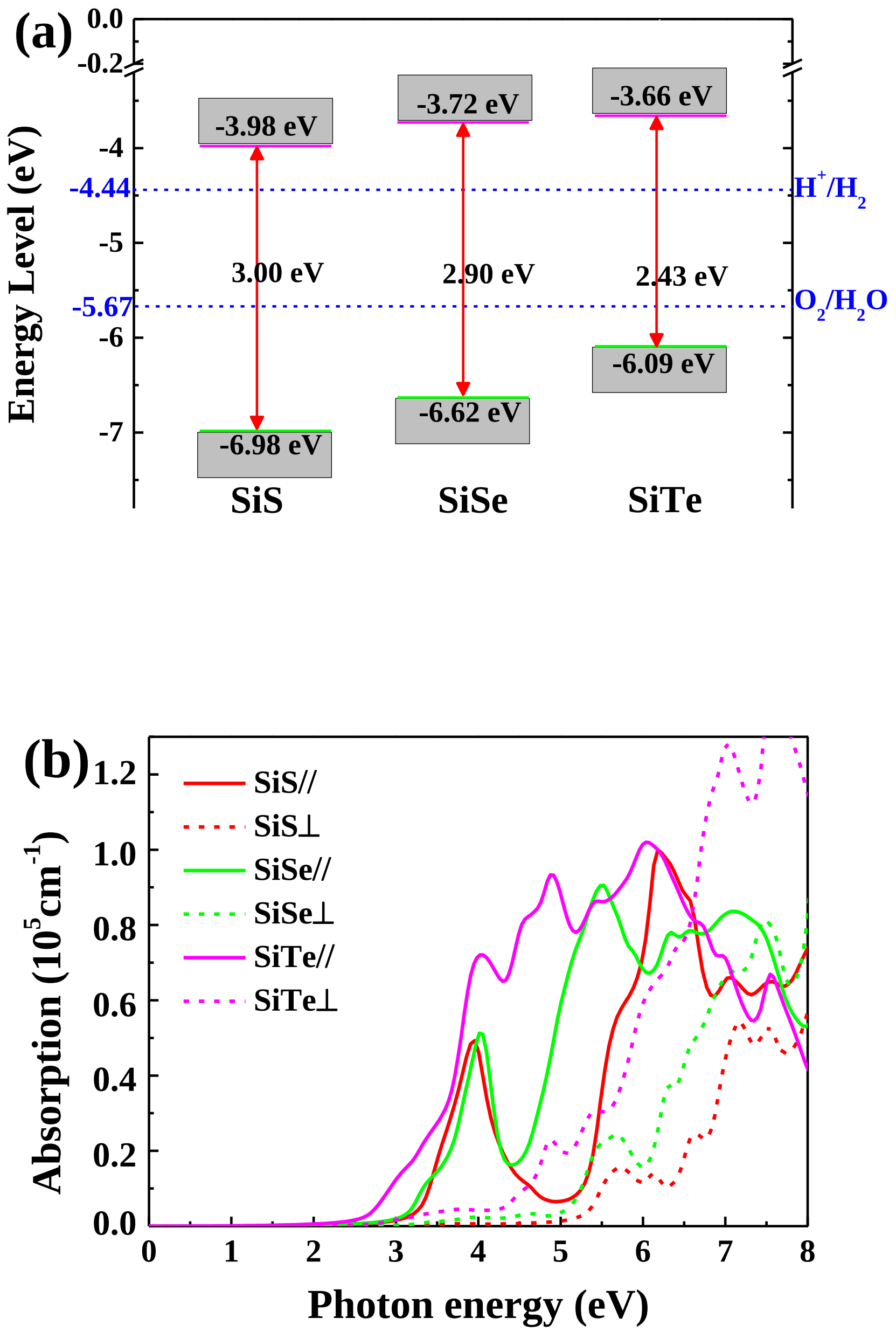

**Fig. 3** (a) VBM and CBM locations of SiX (X=S, Se, Te) monolayers with respect to the vacuum level (labeled as zero energy). The positions of the reduction potential of $H^+$ to $H_2$ and the oxidation potential of $H_2O$ to $O_2$ are indicated by the blue dashed lines. (b)The calculated optical absorption coefficient of SiX monolayers using screened HSE06 hybrid functional.

The band gap results shown in **Table 1** demonstrate that all the SiX monolayers meet the criterion of minimum band gap value (1.23 eV) for water splitting. Yet, another essential condition for photocatalytic water splitting needs also to be fullfilled, that the position of CBM must be higher than the hydrogen reduction potential of $H^+/H_2$ and the VBM must be lower than the oxygen oxidation potential of $H_2O/O_2$. In order to further identify possible photocatalytic activity for SiX monolayers, we calculated the alignment of their CBM/VBM with respect to vacuum level using the HSE06 hybrid functional. **Fig. 3(a)** shows the results as well as the redox potential levels for hydrogen evolution (-4.44 eV) and oxygen evolution (-5.67 eV). It is obvious that the band edge positions of all SiX monolayers perfectly encompass the redox potentials of water at pH = 0, indicating their great applicability in photocatalytic water splitting. Furthermore, to assess the stability of SiX monolayers in aquatic environment, we have also performed AIMD simulations for each SiX monolayer with a water molecular at room temperature. There are neither bond breaking nor new bond formation between $H_2O$ and SiX monolayers (see **Fig. S3**) during the 5 *ps* simulation time at 300 K, suggesting the thermodynamically stability of SiX monolayers in the aquatic environment.

Besides the strict requirement on band edge locations, the ability to harvest solar light is of great importance in practical photocatalytic water splitting. Hence, we further explored the optical properties of SiX monolayers by calculating the absorption spectra in- and out-of-plane using the HSE06 functional. The transverse dielectric function $\varepsilon(\omega)=\varepsilon_1(\omega)+i\varepsilon_2(\omega)$ is used to describe the optical properties of materials,[55,56] where $\omega$ is the photon frequency, $\varepsilon_1(\omega)$ is the real part and $\varepsilon_2(\omega)$ is the imaginary part of the dielectric function, respectively. The absorption coefficient can be evaluated according to the expression[55,56] $\alpha(\omega)=\frac{\sqrt{2}\omega}{c}\left\{\left[\varepsilon_1^2(\omega)+\varepsilon_2^2(\omega)\right]^{\frac{1}{2}}-\varepsilon_1(\omega)\right\}^{\frac{1}{2}}$. As shown in **Fig. 3(b)**, the absorption coefficients of SiX monolayers reach the order of $10^5$ $cm^{-1}$, comparable to that of the organic perovskite solar cells[57]. In-plane absorption is always stronger than that of out-of-plane in the low energy regime, due to the larger cross section area. In addition, the absorption edges for SiS and SiSe monolayers lie in the visible-light region. On the other hand, the SiTe monolayer exhibits excellent light absorption performance in the visible-light region and the absorption onset can be extended to around ~2 eV. The outstanding optical properties pave the way for SiX monolayers to act inphotocatalytic water splitting.

Notwithstanding their appropriate band edge locations with respect to the water splitting levels and the VBM and CBM positions of these SiX monolayers are relatively far from the critical potentials. For instance, the VBM is 1.31 eV lower than the OER reduction potential for SiS monolayer, which may render a low absorption coefficient for solar energy. A promising method to resolve this concern is through bandgap engineering by applying an external mechanical strain, which has been employed for various 2D materials.[19,58,59] Therefore, we further explored the impact of in-plane biaxial strain on the band gaps and band edge positions for SiX monolayers. The biaxial strain was achieved through changing the crystal lattice parameters, and it is mathematically defined as $\varepsilon=(a-a_0)/a_0$, where $a_0$ and $a$ are the equilibrium and strained lattice parameters, respectively. A negative $\varepsilon$ denotes compressive strain, while a positive value refers to tensile strain. **Fig. 4** shows the band gaps and band edge positions of SiX monolayers as a function of biaxial strains. One finds that the band gaps decrease monotonously with either compressive or tensile strains (details shown in **Fig. S4**, **S5** and **S6**). Such band gap evolution stems from the strain-induced electronic energy shifts. As shown in **Fig. 4**, the CBM position is insensitive to the strain, but the VBM is gradually shifted upward upon applying stronger tensile strain for SiX monolayers, resulting in the decrease of band gap. Besides, in SiTe monolayer the VBM moves upward with the compressive strain. It is also noted that for SiS and SiSe monolayers, the CBM and VBM levels are approaching the optimal locations for photocatalytic water splitting, under various biaxial strain ranges considered (at pH=0, see **Fig. 4(a) and 4(b)**). For SiTe monolayer, when the strain ranges from -3% to 8%, high-efficiency water-splitting is expected as well (at pH=0, see **Fig. 4(c)**). In addition, adjusting the pH value[19,49,60] was another possible solution to tune the water-splitting properties of SiX monolayers, since the redox potentials for water increase with pH by pH×$k_B T$ln10.[61] In fact, photocatalytic water-splitting usually occurs in a neutral environment. Encouragingly, the redox potentials of photocatalytic water-splitting are still located inside the energy gap of the strained SiX monolayers (*i.e.* 0—15% for SiS, -4%—9% for SiSe and -3%—8% for SiTe) at pH=7. Compared with SiS and SiSe monolayers, the SiTe monolayer would present an enhanced oxygen evolution activity in the neutral environment as its VBM is much closer to the redox potential.

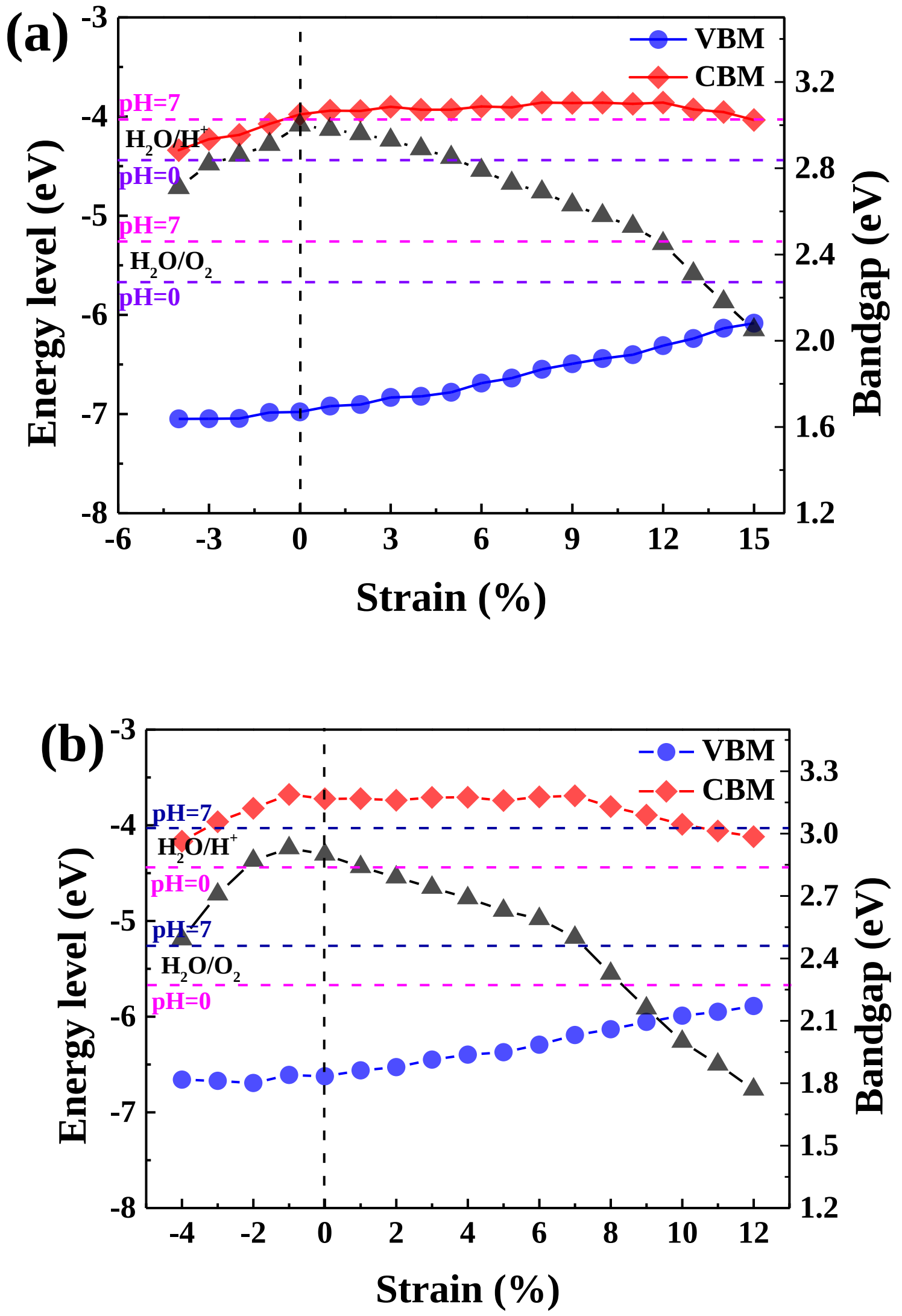

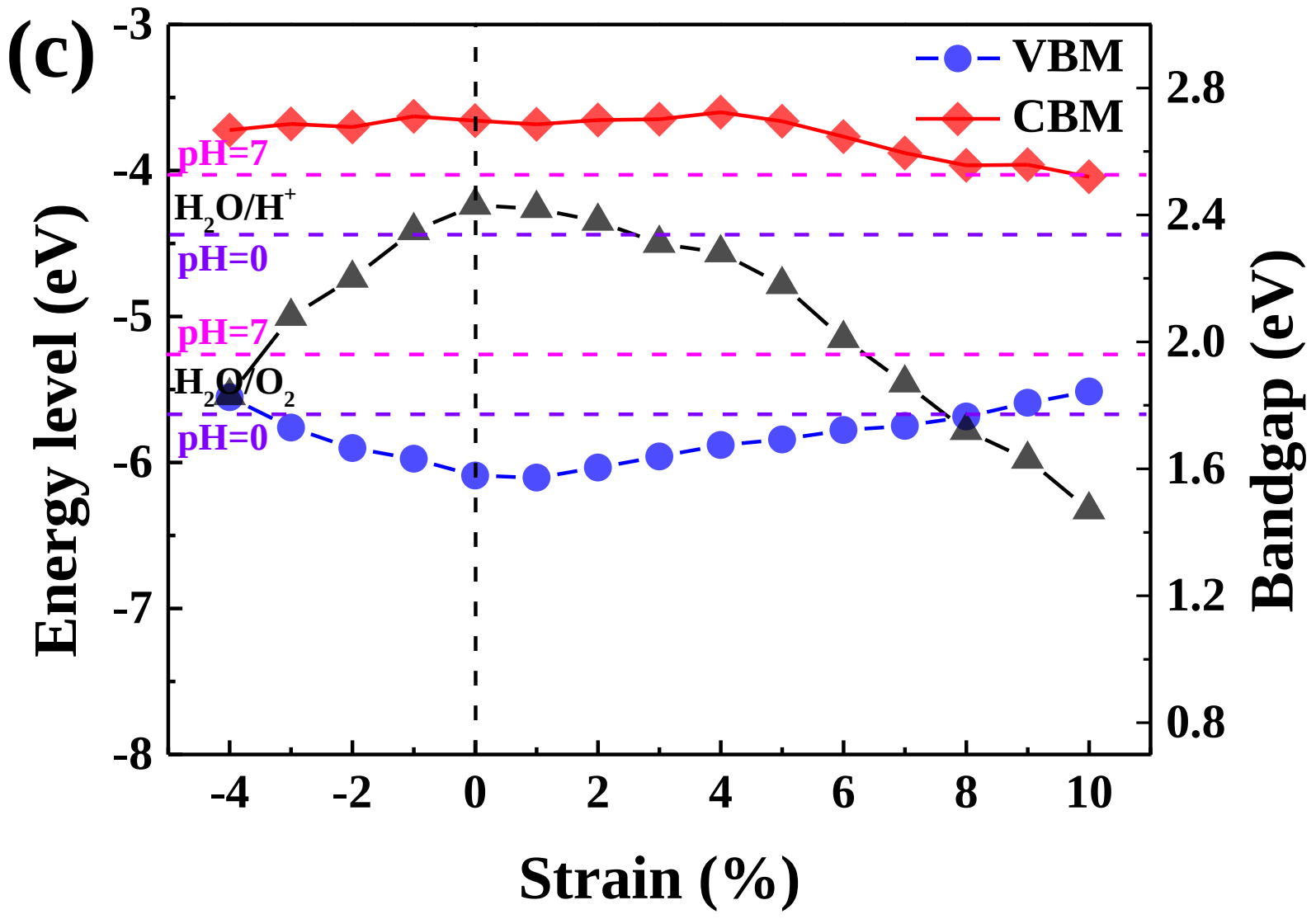

**Fig. 4** Strain effects on band gap and band-edge positions of the (a) SiS; (b) SiSe; (c) SiTe monolayers, respectively. Obtained from HSE06 calculations. The redox potentials of water splitting at pH = 0 (blue dashed line) and pH = 7 (wine dashed line) are shown for comparison.

Another route to adjust the band gaps and band edge locations is by increasing the number of layers in 2D materials, which in general causes drastic changes to their electronic structures.[34,62,63] Therefore, bilayer SiX were also considered in our work, where three different stacking configurations of SiX were considered (see **Fig. S7**) and the energy differences between various configurations are listed in **Table S2**. The AA stacking, where the top layer resides on the bottom layer directly, was identified as the most stable stacking configuration for SiS and SiSe. Yet, the AB stacking configuration where the Si atoms of the top layer are located on the hexagonal hollow sites of the bottom layer, was found to be the most energetically favorable for SiTe. Only these most stable stacking configurations were considered in further calculations. **Fig. S8** shows the band structures of bilayer SiX obtained using the HSE06 functional. As expected, the band gaps of bilayer SiX (2.19 eV, 1.72 eV, and 0.40eV for SiS, SiSe, and SiTe, respectively) are lower than their monolayer counterparts. The redox energy levels of bilayer SiX with respect to the oxygen oxidation and hydrogen reduction potential levels have been compared to evaluate their photocatalytic performance, as shown in **Fig. 5(a)**. One clearly finds that bilayer SiSe is more favorable for water splitting than its monolayer counterparts, but bilayer SiTe is no longer suitable for water splitting. The case of bilayer SiS is more complicated since its band gap is at a suitable value, but the CBM over hydrogen reduction potential of SiS is too small to be sufficient for $H_2$ production. It needs to be point out that the band edge alignment for bilayer SiS becomes suitable for overall water splitting by adding an extra tensile force of 3% **(Fig. S9)**. The optical absorption properties of SiS and SiSe are analyzed in **Fig. 5(b)**. Remarkably, there is a great improvement in the absorption coefficients of SiX which can reach $10^5$ $cm^{-1}$, comparable to the organic perovskite solar cells.[64] Considering both band edge location and solar light absorption, the bilayer SiSe is an even better candidate than those monolayer SiX.

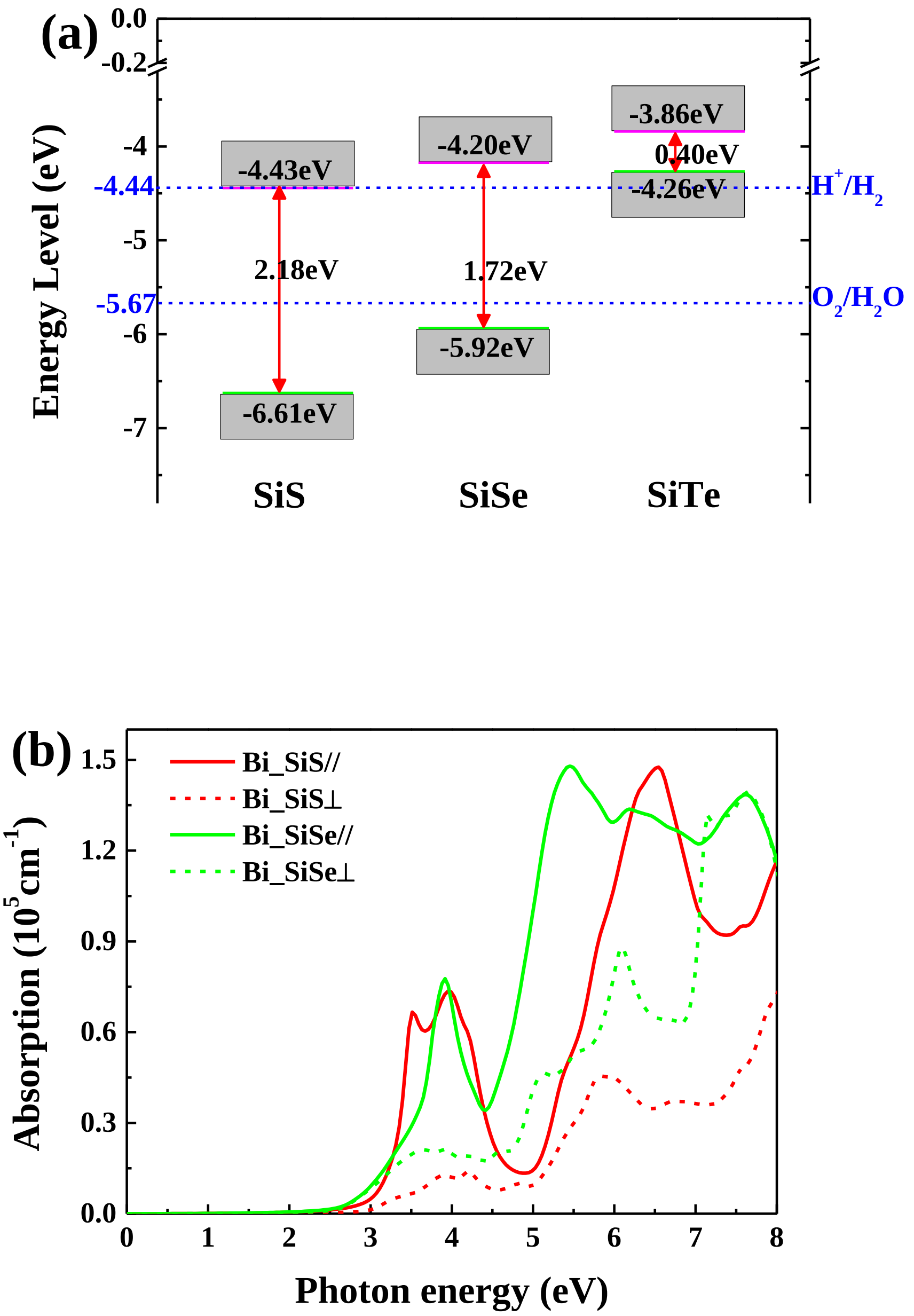

**Fig. 5** (a) VBM and CBM locations of SiX (X=S, Se, Te) bilayers with respect to the vacuum level (labeled as zero energy). The positions of the reduction potential of $H^+$ to $H_2$ and the oxidation potential of $H_2O$ to $O_2$ are indicated by the blue dashed lines. (b)The calculated optical absorption coefficients of SiX bilayers, calculated using the screened HSE06 hybrid functional.

**Table 2.** Calculated effective mass $m^*$ ($m_e$), deformation potential constant $|E_1^i|$ (eV), elastic modulus $C_{2D}$ (N m$^{-1}$), carrier mobility $\mu_{2D}$ (cm$^2$ V$^{-1}$s$^{-1}$) for monolayer and bilayer SiX (ML SiX and BL SiX for short) along the *a* and *b* directions.

| Material | Carrier type | $m_a^*$ | $m_b^*$ | $\|E_{1a}\|$ | $\|E_{1b}\|$ | $C_a^{2D}$ | $C_b^{2D}$ | $\mu_a^{2D}$ | $\mu_b^{2D}$ |
|---|---|---|---|---|---|---|---|---|---|
| ML SiS | electron | 1.93 | 1.93 | 5.42 | 3.15 | 50.66 | 51.02 | 9.87 | 29.35 |
| | hole | 1.34 | 1.45 | 0.77 | 1.00 | 50.66 | 51.02 | 973.66 | 540.23 |
| ML SiSe | electron | 0.73 | 0.56 | 4.76 | 6.40 | 41.39 | 41.36 | 83.36 | 59.23 |
| | hole | 1.02 | 1.18 | 5.94 | 0.70 | 41.39 | 41.36 | 22.37 | 1397.00 |
| ML SiTe | electron | 0.61 | 0.48 | 4.47 | 5.72 | 34.29 | 34.32 | 109.66 | 86.48 |
| | hole | 0.24 | 0.46 | 4.13 | 4.82 | 34.29 | 34.32 | 540.88 | 206.01 |
| BL SiS | electron | 0.36 | 0.37 | 2.69 | 2.76 | 99.22 | 99.26 | 2223.49 | 2040.29 |
| | hole | 1.44 | 1.33 | 1.95 | 0.62 | 99.22 | 99.26 | 280.00 | 3017.85 |
| BL SiSe | electron | 0.08 | 0.05 | 2.15 | 1.39 | 86.07 | 86.04 | 89684.12 | 340186.75 |
| | hole | 1.52 | 1.39 | 2.26 | 1.59 | 86.07 | 86.04 | 163.30 | 360.79 |

Last but not least, high carrier mobilities are also highly desired to suppress the unfavorable electron-hole recombination in photocatalytic water splitting. Thus, we systematically calculated the carrier mobility (electrons and holes) based on the deformation potential theory proposed by Bardeen and Shockley.[65] The carrier mobility of 2D materials can be calculated by the following equation [66–68]

$$\mu_{2D} = \frac{e\hbar^3 C_{2D}}{k_B T m^* m_d (E_1^i)^2},$$

where $\hbar$ is the reduced Planck constant, $k_B$ is the Boltzmann constant, $m^*$ is the effective mass in the direction of transport, $m_d$ is the average effective mass determined by $m_d = (m_a^* m_b^*)^{1/2}$, and $T$ is the temperature ($T$=300K). The elastic modulus $C_{2D}$ of the longitudinal strain in the propagation direction is derived from $(E-E_0)/S_0=C_{2D}(\Delta l/l_0)^2/2$, where $E$ is the total energy of the 2D structure, and $S_0$ is the lattice area of the equilibrium supercell. The deformation potential constant $E_1^i$ is defined as $E_1^i=\Delta E_i/(\Delta l/l_0)$. Here $\Delta E_i$ is the energy change of the $i^{th}$ band under proper cell compression and dilatation (calculated using a step of 0.5%), $l_0$ is the lattice constant in the transport direction and $\Delta l$ is the deformation of $l_0$. We utilized a rectangle supercell of SiX monolayers and bilayers to differentiate the carrier conduction along the *a* and *b* directions (see **Fig. S10-S15**).

As summarized in **Table 2**, the elastic moduli are isotropic for monolayer SiX, and the values are slightly higher than that of GeTe monolayer ($C_{2D}$ = 29.25 N $m^{-1}$). [49] For SiS monolayer, the effective masses of both electron and hole along the *a*/*b* directions are larger than those of SiSe and SiTe monolayers, stemming from the flat band feature in SiS. In addition, compared with SiS, the effective masses of SiSe and SiTe monolayers show stronger anisotropy. The deformation potentials $E_l$ of SiX monolayers also exhibit anisotropy, and that of holes is generally smaller than that of electrons (expect for SiSe monolayer along the *a* direction). This contributes to the relatively large hole mobilities for SiX monolayers, *i.e.* 974 $cm^2$ $V^{-1}$ $s^{-1}$ for SiS monolayer, 1397 $cm^2$ $V^{-1}$ $s^{-1}$ for SiSe monolayer, and 540 $cm^2$ $V^{-1}$ $s^{-1}$ for SiTe monolayer. Note that the electron mobilities of SiX monolayers are rather low, comparing with that of holes. For instance, the calculated hole mobility of SiS monolayer along the *a* direction is about 98 times higher than the electron mobility for the same direction (9.87 $cm^2\ V^{-1}\ s^{-1}$), exhibiting ultra-high anisotropy. Such unbalanced carrier mobility of SiX monolayers would significantly hinder the recombination of photo-generated electrons and holes, favoring long-term photocatalytic activity. We also note that the hole carrier mobilities of SiX monolayers are larger than the theoretical values of many other proposed 2D catalysts for water splitting, such as $MoS_2$ (~200 $cm^2\ V^{-1}\ s^{-1}$) [69] and PdSP (~312 $cm^2\ V^{-1}\ s^{-1}$) [70], indicating a rather fast redox reaction. For bilayer SiS/Se, the elastic moduli are isotropic like their monolayer counterparts, but the values are two times that of the monolayers. The electron effective masses of the SiS bilayer along the *a* and *b* directions are much smaller than its monolayer counterpart, but the effective masses of holes are nearly the same for bilayer and monolayer SiS. Combined with a relatively small deformation potential, bilayer SiS exhibits much larger electron mobilities (2223 $cm^2\ V^{-1}\ s^{-1}$ and 2040 $cm^2\ V^{-1}\ s^{-1}$) than those of monolayer SiS. And hole mobilities with high anisotropy have also been revealed in the bilayer SiS, due to the strong anisotropy of deformation potentials. Notably, bilayer SiSe has very tiny effective electron masses along the *a* and *b* directions, due to its steep band feature (**Fig. S14a**). Under the joint action of tiny electron effective masses and small deformation potential, bipolar SiSe exhibits ultrahigh electron mobilities along both *a* and *b* directions (89684 $cm^2$ $V^{-1}$ $s^{-1}$ and 340187 $cm^2$ $V^{-1}$ $s^{-1}$, respectively). Concerning the holes, however, the carrier mobility is only at the level of few hundred $cm^2$ $V^{-1}$ $s^{-1}$, demonstrating ultrahigh discrepancy (~248 times). All of these excellent properties reveal that bilayer SiS and SiSe are of great potential in the field of photocatalytic water splitting.

## Conclusion

To summarize, we have shown that the 2D semiconducting silicon chalcogenides, *i.e.* SiX (X=S, Se and Te) monolayers, are a series of remarkable candidates for high performance photocatalytic water splitting. SiX monolayers are dynamically, mechanically and thermodynamically stable at room temperature. They possess moderate band gaps around 2.43 eV~3.00 eV, which can be manipulated by applying biaxial strains or increasing the number of layers. They are also featured with considerable carrier mobilities and substantial light absorption in the range of the solar spectra. We show that the band alignments of monolayer SiX and bilayer SiSe well fit the requirement of photocatalytic water splitting, and they may serve as effective photocatalysts.

## Conflicts of interest

There are no conflicts to declare.

## Acknowledgement

This work was supported by "The National Key Research and Development Program of China (17YFB0405601)", and the National Natural Science Foundation of China under Grant No. 11704134. K.-H. Xue received support from China Scholarship Council (No. 201806165012).

Support Information for

# Two dimensional silicon chalcogenides with high carrier mobility for photocatalytic water splitting

Yun-Lai Zhu,[1+] Jun-Hui Yuan,[1+] Ya-Qian Song,[1] Sheng Wang,[1] Kan-Hao Xue,[1,2*] Ming Xu,[1] Xiao-Min Cheng,[1*] Xiang-Shui Miao[1]

[1]Wuhan National Research Center for Optoelectronics, School of Optical and Electronic Information, Huazhong University of Science and Technology, Wuhan 430074, China

[2]IMEP-LAHC, Grenoble INP–Minatec, 3 Parvis Louis Néel, 38016 Grenoble Cedex 1, France

[+]The authors Y.-L. Zhu and J.-H. Yuan contributed equally to this work.

**Corresponding Authors**

*E-mail: xmcheng@hust.edu.cn (X.-M. Cheng)  xkh@hust.edu.cn (K.-H. Xue)

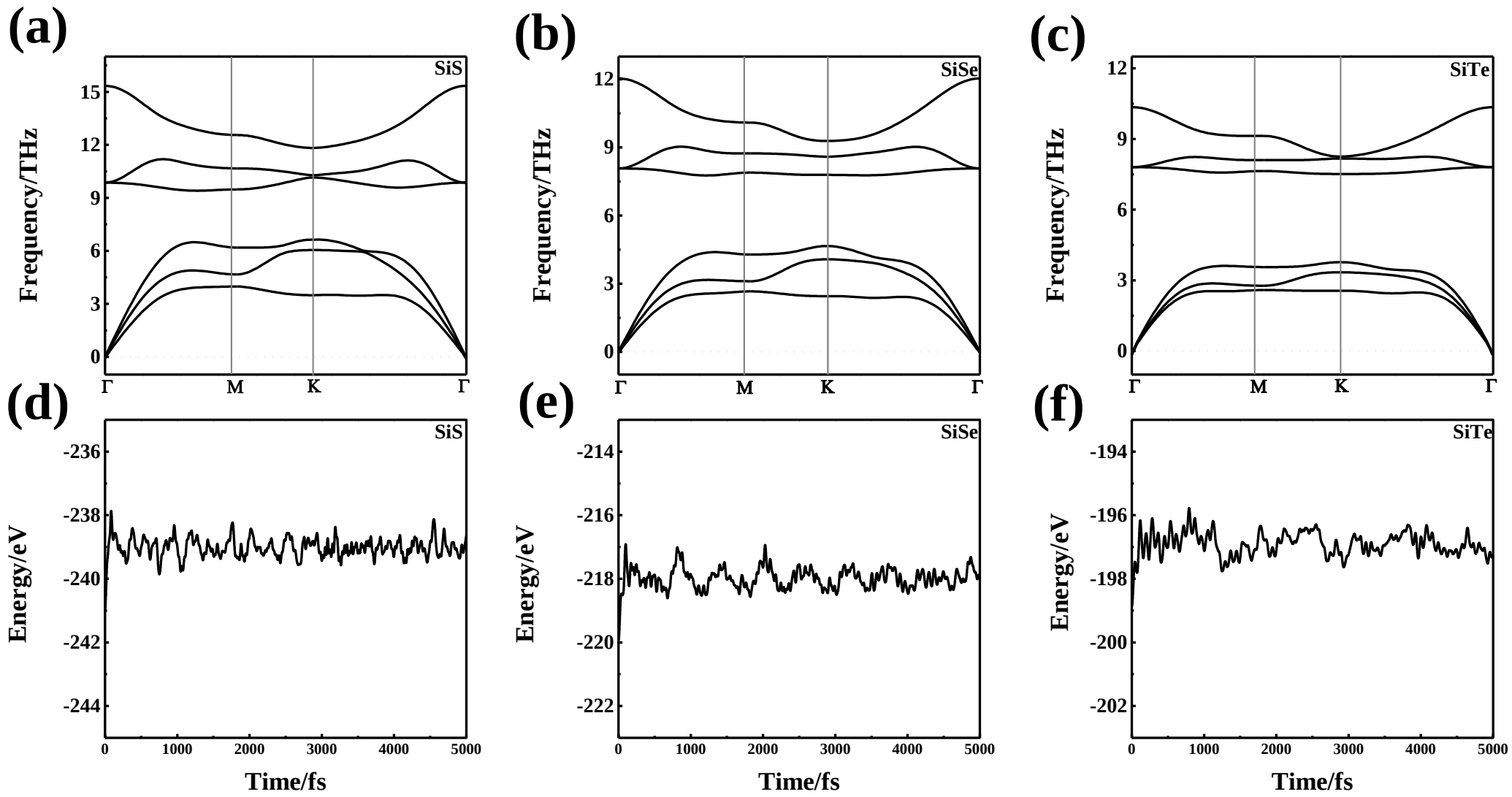

**Fig. S1** Phonon band structures of (a) SiS, (b) SiSe, and (c) SiTe monolayers, along the lines between high-symmetric *k* points in the Brillouin zone. The total energy fluctuations during AIMD simulations

are shown for (d) SiS, (e) SiSe, and (f) SiTe monolayers at 300K.

**Table S1** Effective independent elastic constants ($C_{ij}$, in N m$^{-1}$), Young's moduli ($Y$, in N m$^{-1}$) and Poisson's ratios ($v$) of SiX (X = S, Se, Te) monolayers.

| | $C_{11}$ | $C_{12}$ | $C_{22}$ | $C_{44}$ | $Y$ | $v$ |
|---|---|---|---|---|---|---|
| SiS | 52.20 | 52.20 | 6.96 | 22.63 | 51.27 | 0.13 |
| SiSe | 42.71 | 42.71 | 7.21 | 17.74 | 41.49 | 0.17 |
| SiTe | 35.54 | 35.54 | 6.44 | 14.55 | 34.37 | 0.18 |

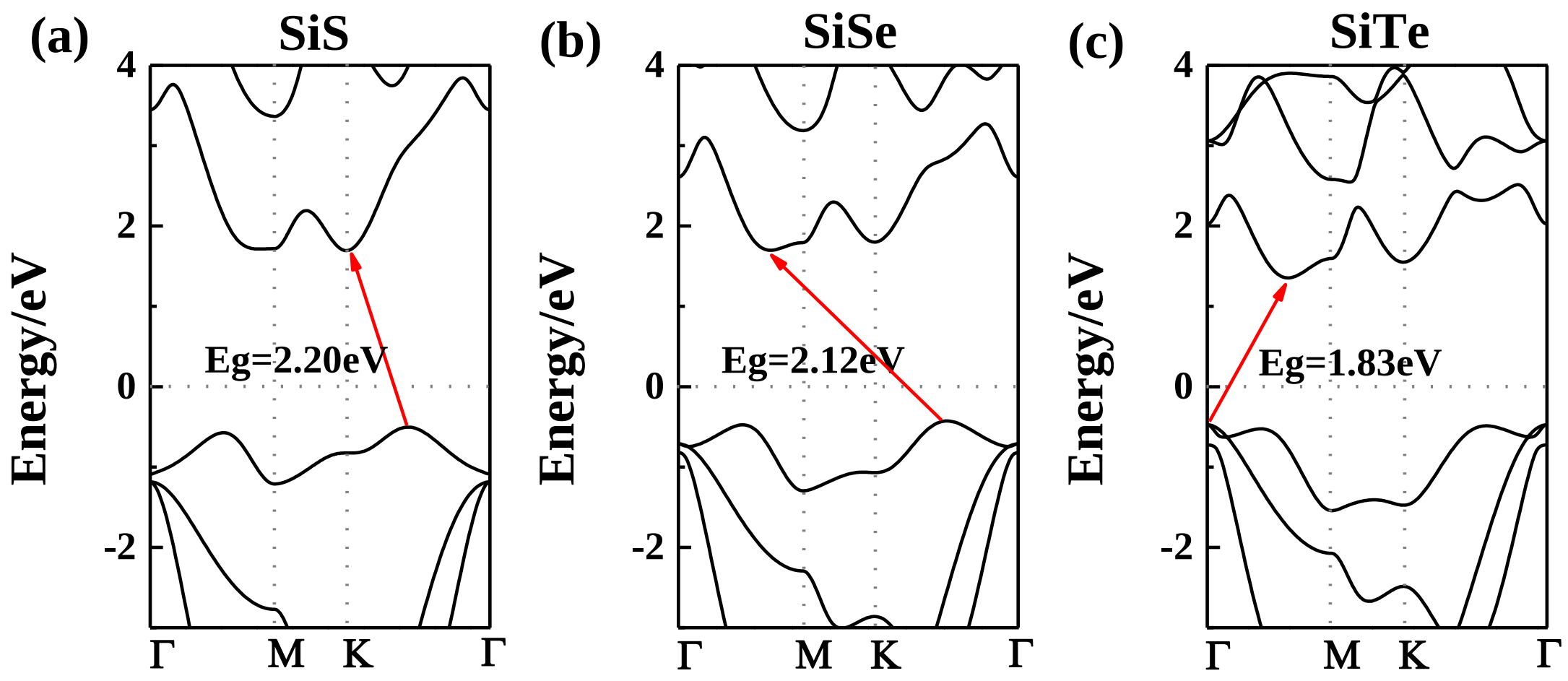

**Fig. S2** Electronic band structures of the (a) SiS, (b) SiSe, and (c) SiTe monolayers, calculated with the PBE functional.

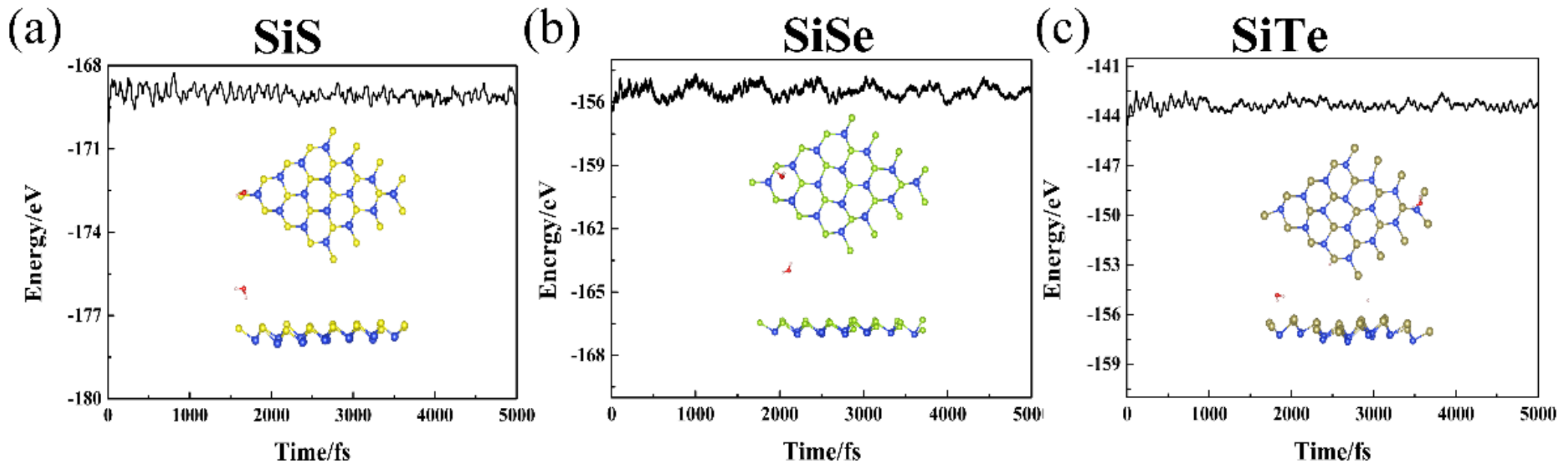

**Fig. S3** The vibration of the total energies for (a) SiS, (b) SiSe, and (c) SiTe monolayers during *ab initio* molecular dynamics simulations in the water environment at 300K. The insets show snapshots of the structures after the 5 *ps* simulation time from the 4 × 4 supercell with a water molecule.

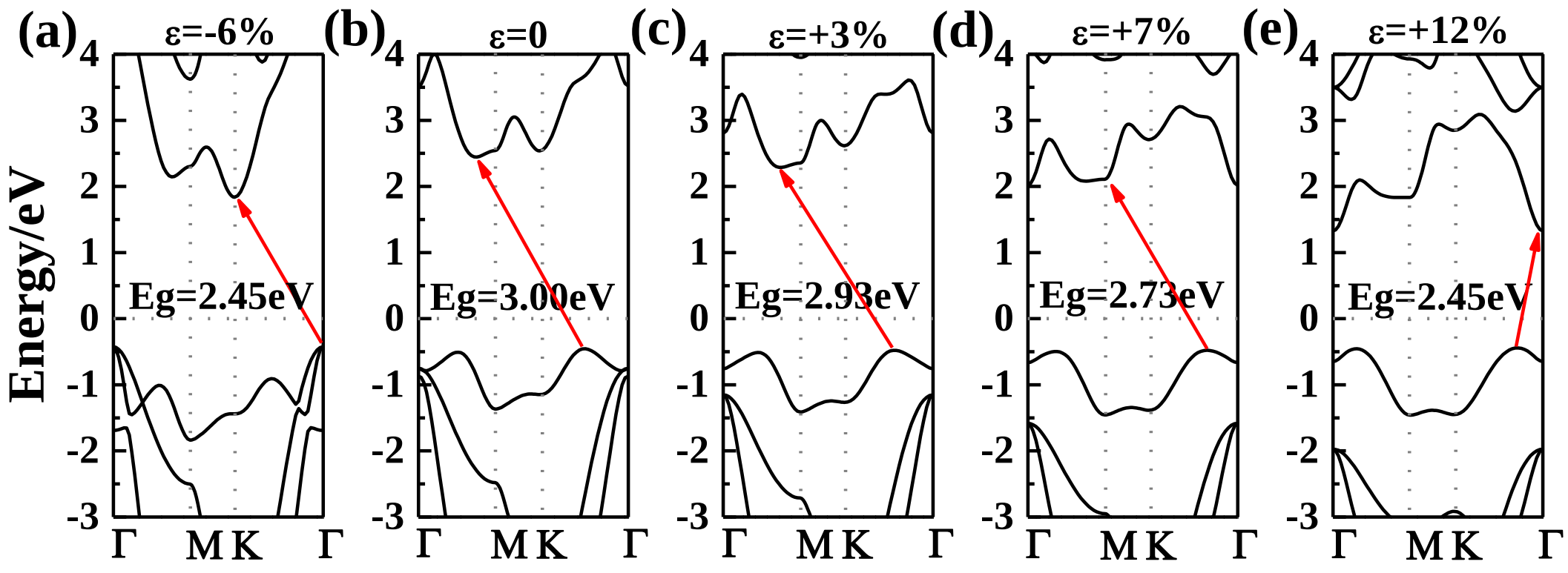

**Fig. S4** Electronic band structures of the SiS monolayer under (a) -6%; (b) 0; (c) 3%; (d) 7%; (e) 12% biaxial strains, respectively, calculated with the HSE06 functional. The Fermi energy is set to zero.

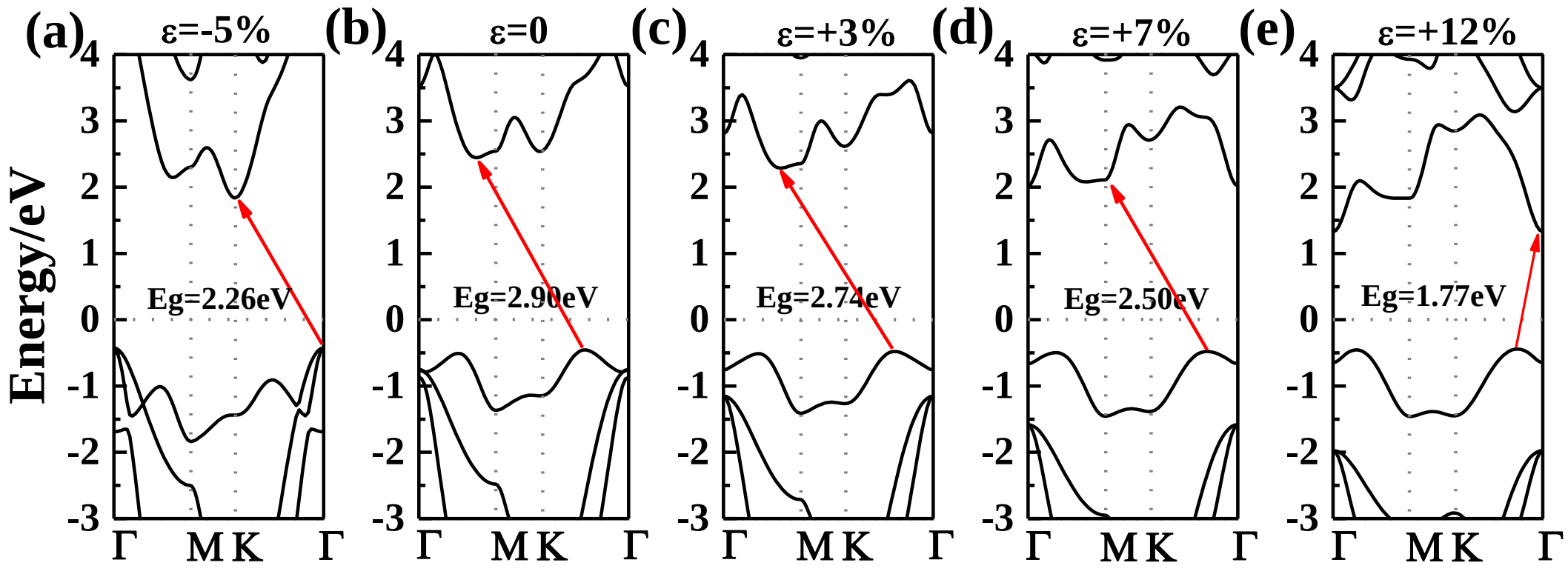

**Fig. S5** Electronic band structures of the SiSe monolayer under (a) -5%; (b) 0; (c) 3%; (d) 7%; (e) 12% biaxial strains, respectively, calculated with the HSE06 functional. The Fermi energy is set to zero.

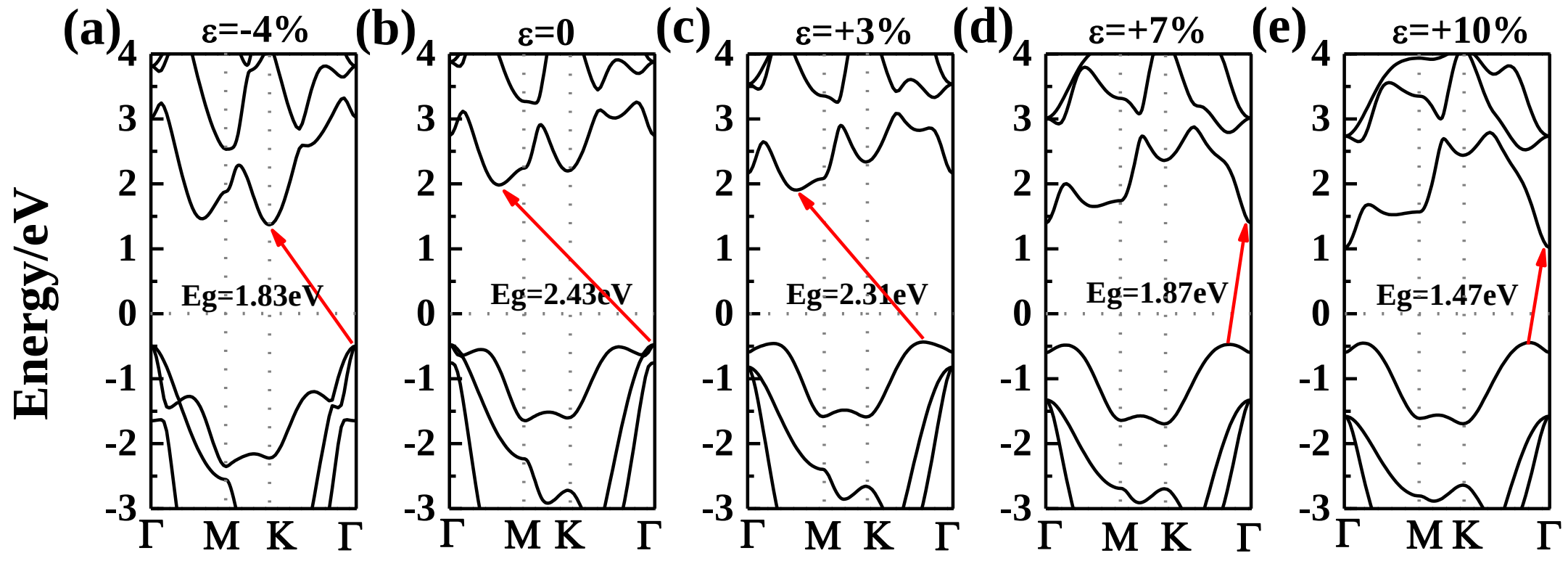

**Fig. S6** Electronic band structures of the SiTe monolayer under (a) -4%; (b) 0; (c) 3%; (d) 7%; (e) 10% biaxial strains, respectively, calculated with the HSE06 functional.

The evolution of band structures of the SiS monolayer with different levels of strain are shown in Fig. S4. Here, several representative band structure characteristics of SiS are chosen in each region. It is clearly observed that the CBM shifts from the K point to the M point, and VBM still lies along the K-Γ line when subjected to a biaxial tensile strain of 3%. With the enhancement of biaxial strain up to 12%, the CBM shifts from the M point to the Γ point, and VBM is kept the same. However, when the SiS monolayer is subjected to a compressed strain of -6%, the CBM remains unchanged while VBM shifts to the Γ point from the location along K-Γ. Those phenomena were also found in SiSe and SiTe monolayers with different levels of strain, as shown in Fig. S5-S6, which account for the band gap narrowing.

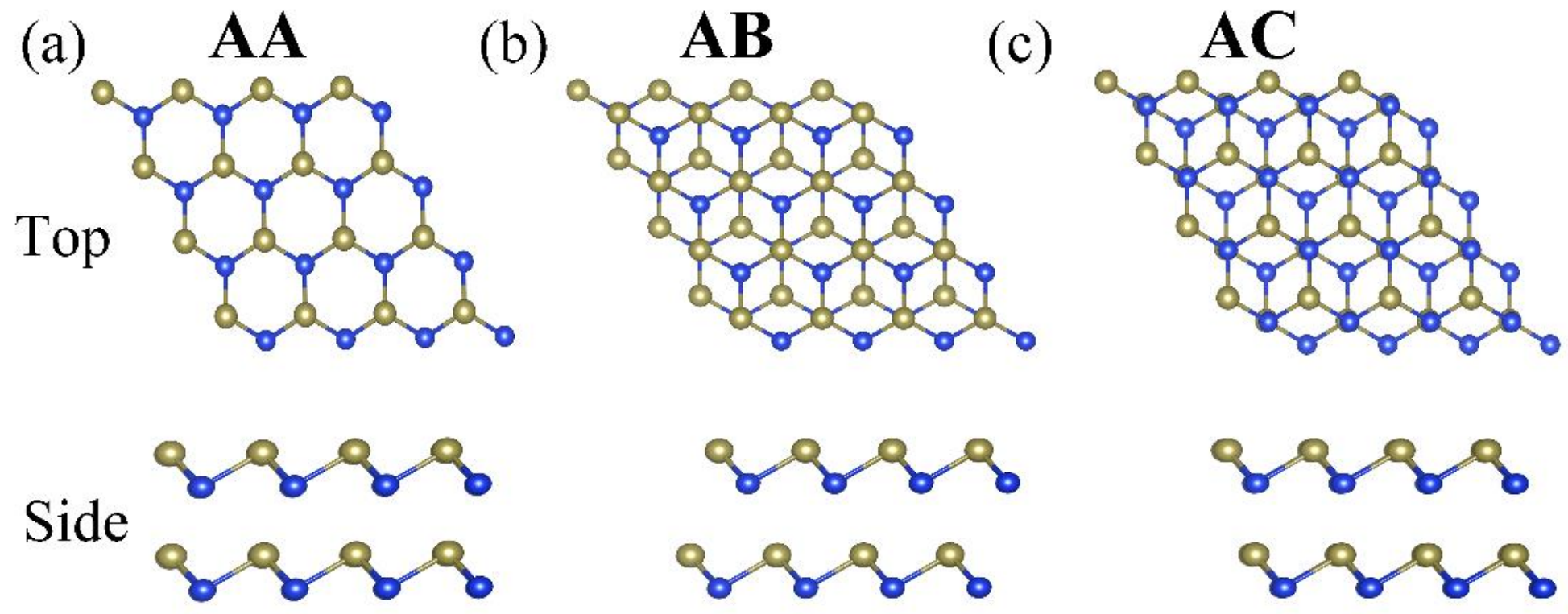

**Fig. S7** Demonstration of the three stacking configurations of (a) AA; (b) AB; (c) AC of bilayer SiX (X=S, Se, Te).

**Table S2**. Energy difference Δ$E$ (unit: meV) between various stacking configurations of SiX (X=S, Se, Te). The most energetically favorable configuration is set to zero energy for each material.

| | SiS | SiSe | SiTe |
|---|---|---|---|
| AA | 0 | 0 | 129 |
| AB | 20.4 | 26.8 | 0 |
| AC | 12.8 | 0.829 | 130 |

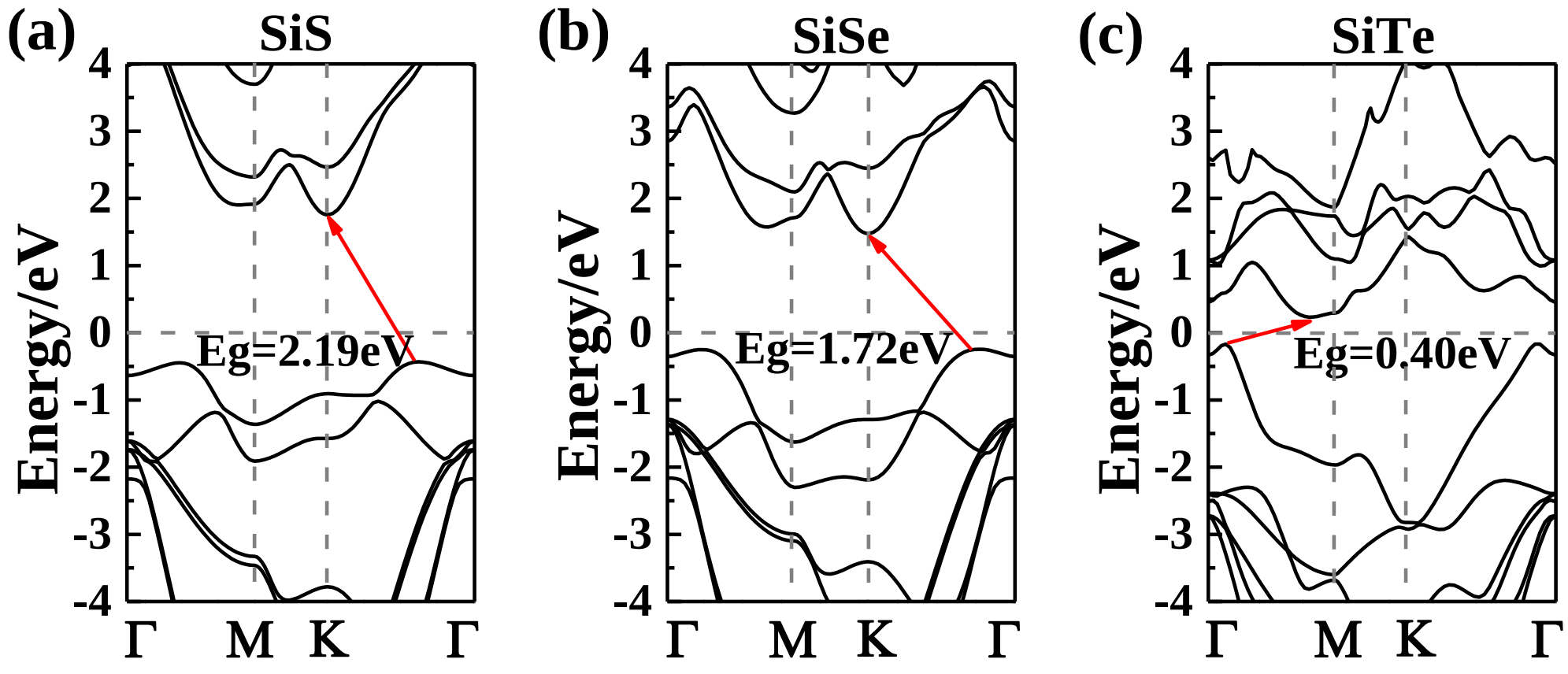

**Fig. S8** Electronic band structures of bilayer SiX: (a) bilayer SiS; (b) bilayer SiSe with AA stacking; (c) bilayer SiTe with AB stacking.

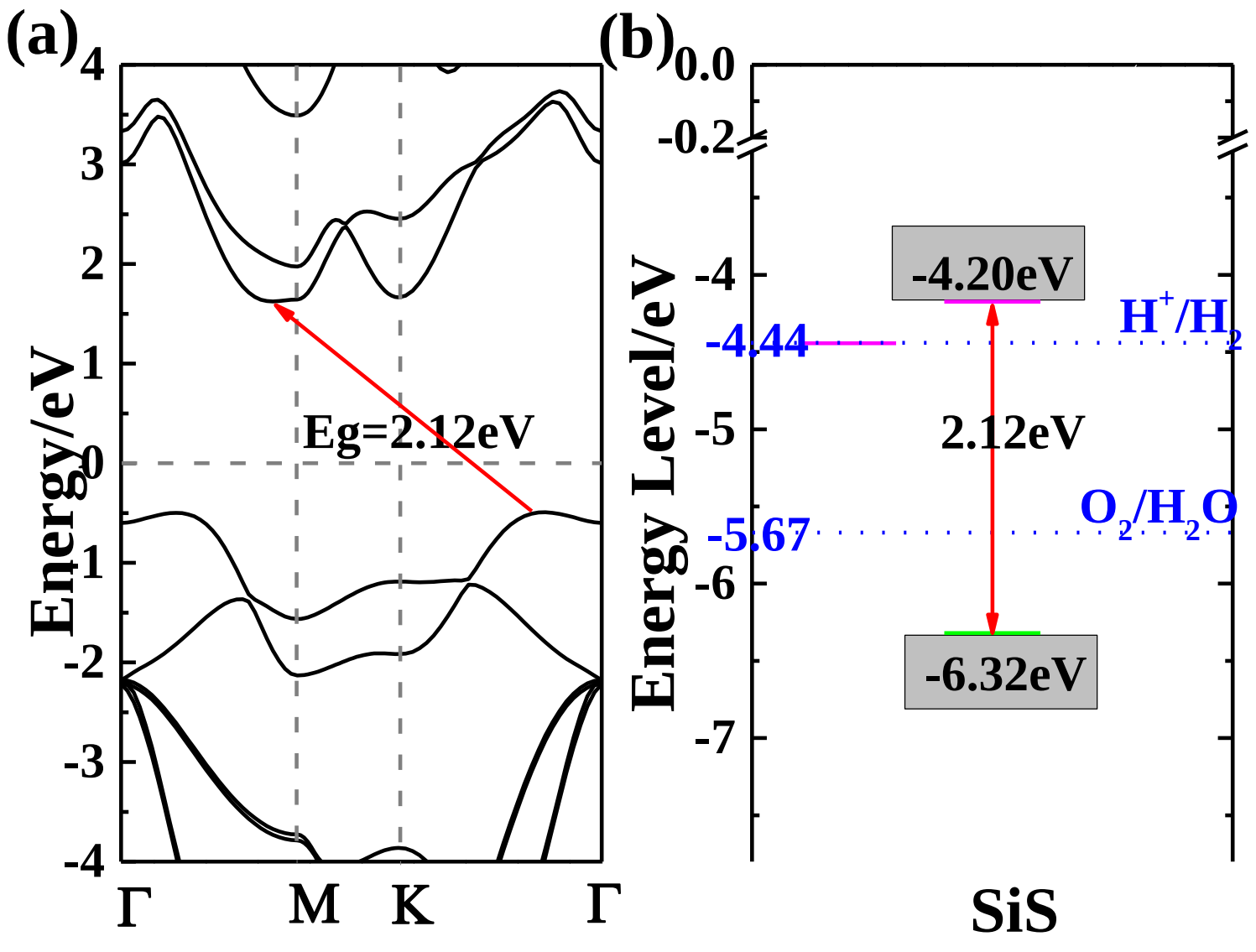

**Fig. S9** (a) The electronic band structure of bilayer SiS with 3% of tensile strain; (b) The corresponding band alignments.

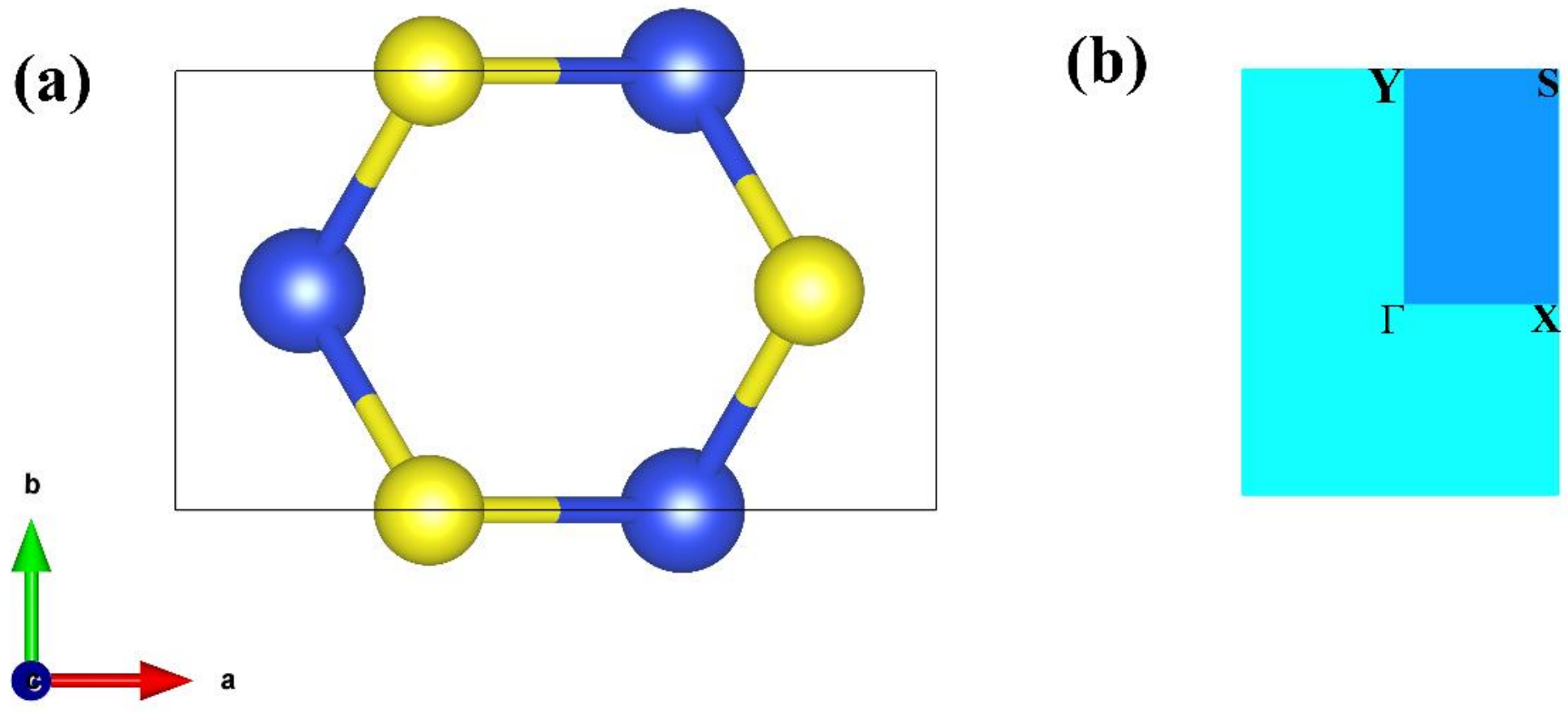

**Fig. S10** (a) Crystal structure and (b) the corresponding first Brillouin zone with high symmetry points of SiX monolayers (bilayers) in an orthogonal supercell.

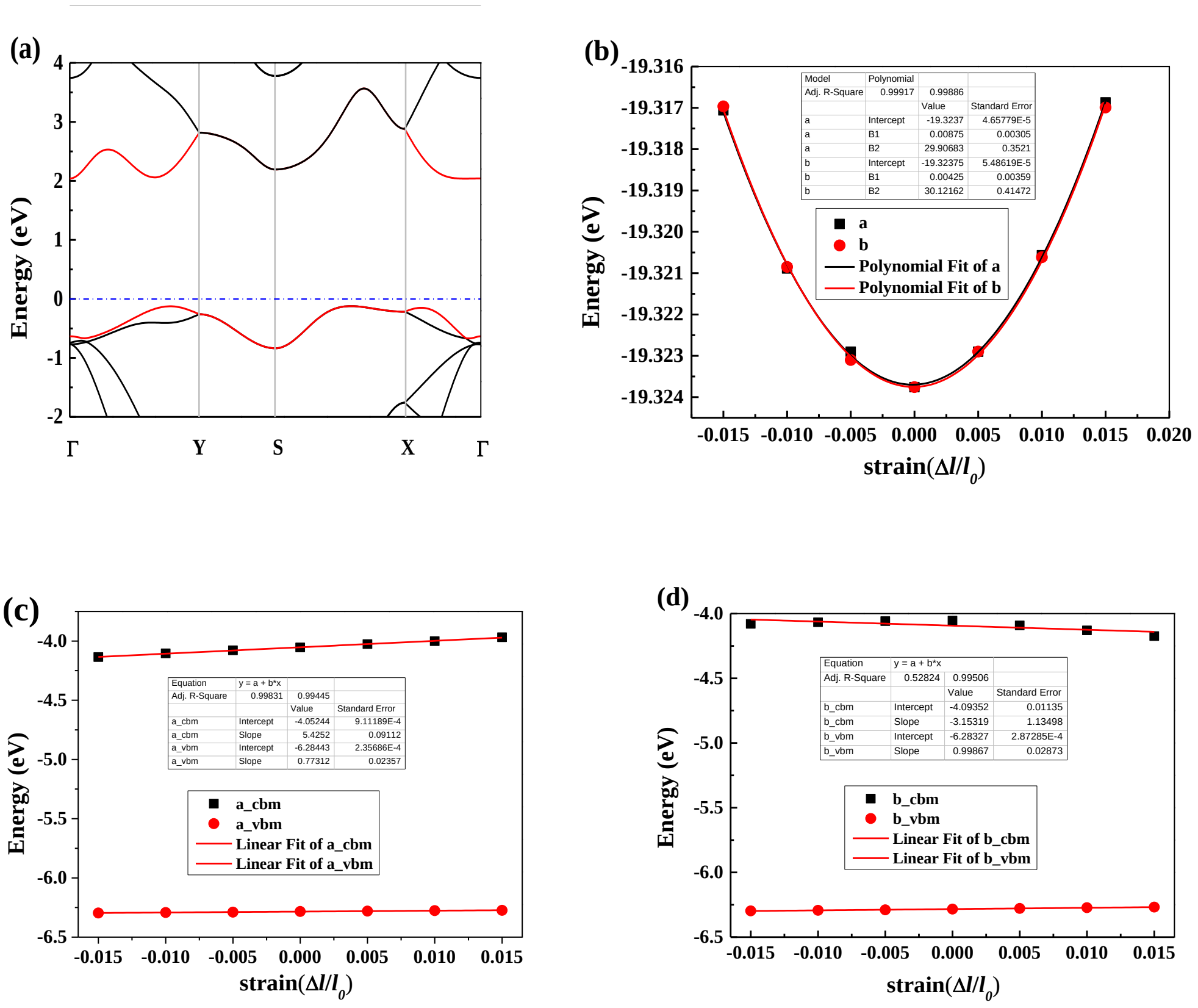

**Fig. S11** (a) Electronic band structure of SiS monolayer in an orthogonal supercell; (b) Total energy difference between the unstrained and strained SiS monolayers along the *a* and *b* directions; (c) Energy shift of VBM and CBM for monolayer SiS with respect to the lattice dilation and compression along the *a* direction; (d) Energy shift of VBM and CBM for monolayer SiS with respect to the lattice dilation and compression along the *b* direction.

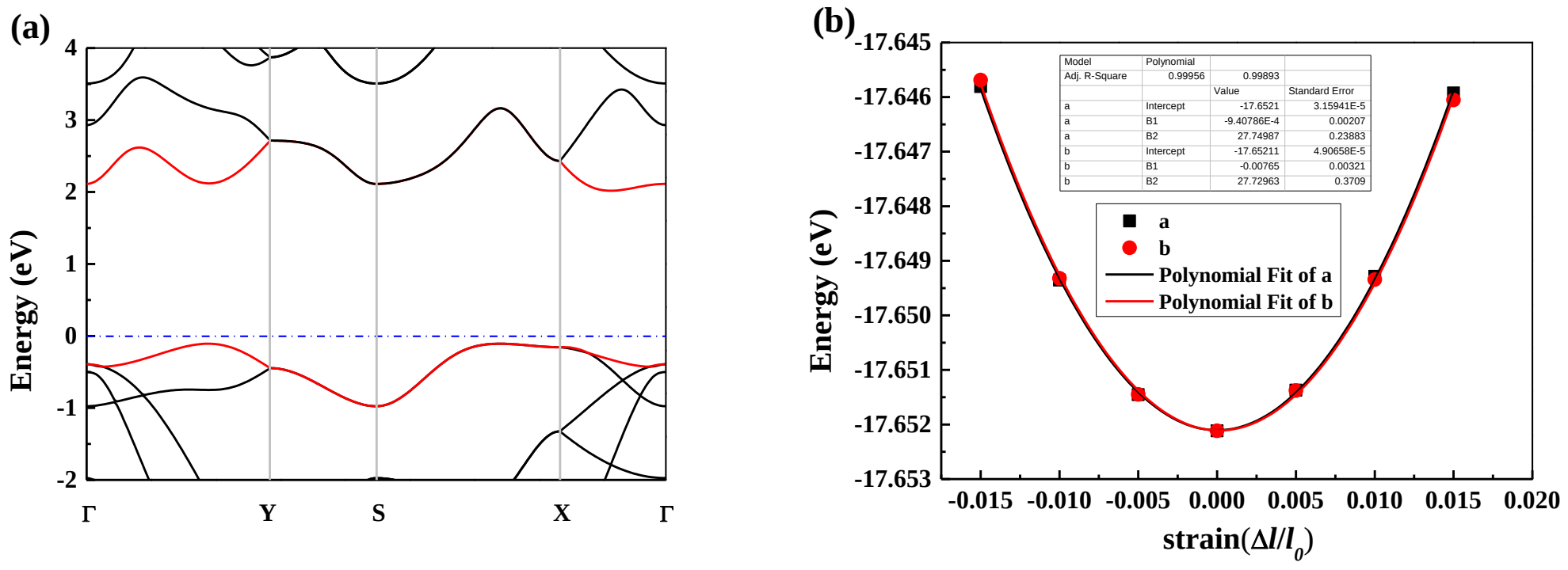

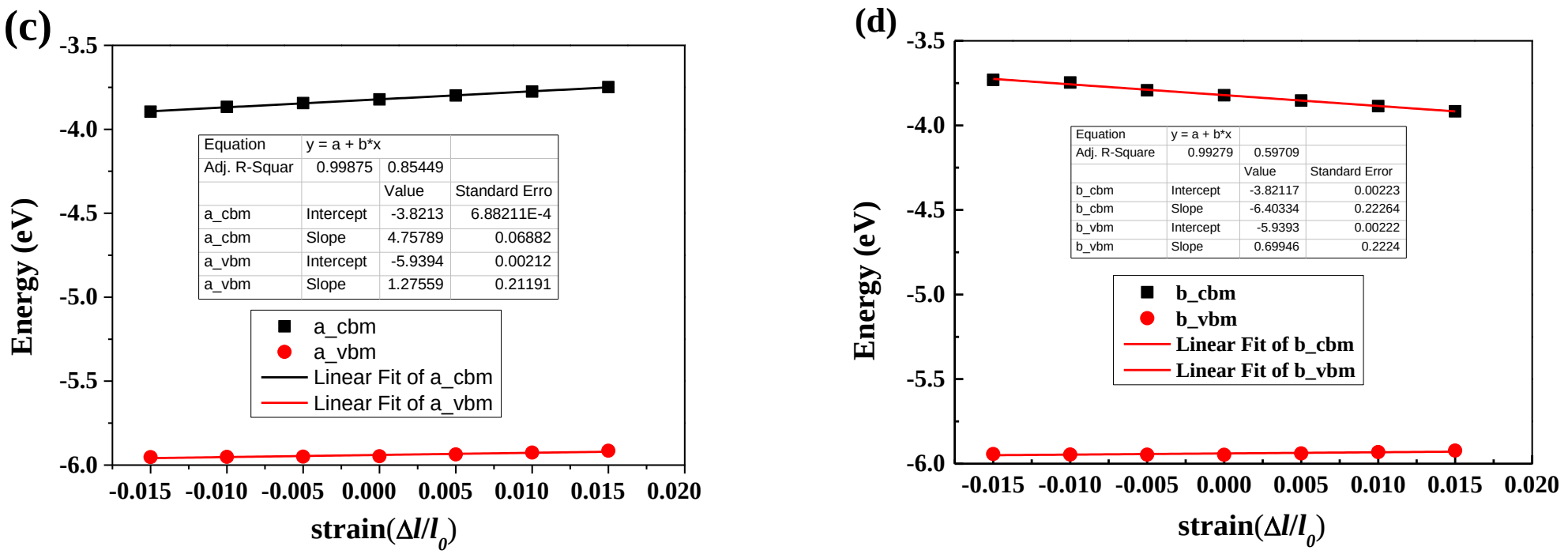

**Fig. S12** (a) Electronic band structure of SiSe monolayer in an orthogonal supercell; (b) Total energy difference between the unstrained and strained SiSe monolayers along the *a* and *b* directions; (c) Energy shift of VBM and CBM for monolayer SiSe with respect to the lattice dilation and compression along the *a* direction; (d) Energy shift of VBM and CBM for monolayer SiSe with respect to the lattice dilation and compression along the *b* direction.

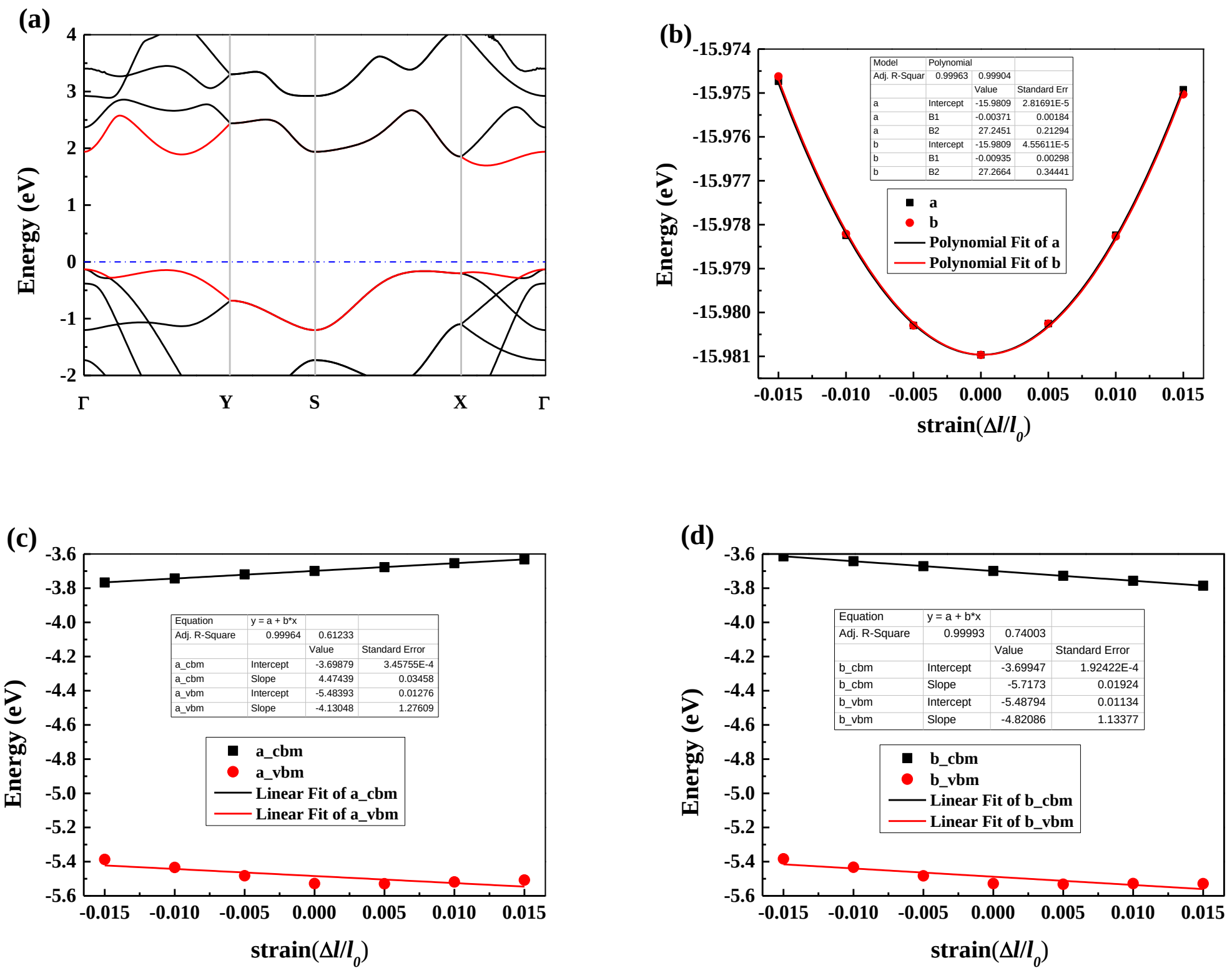

**Fig. S13** (a) Electronic band structure of SiTe monolayer in an orthogonal supercell; (b) Total energy difference between unstrained and strained SiTe monolayers along the *a* and *b* directions; (c) Energy shift of VBM and CBM for monolayer SiTe with respect to the lattice dilation and compression along the *a* direction; (d) Energy shift of VBM and CBM for monolayer SiTe with respect to the lattice dilation and compression along the *b* direction.

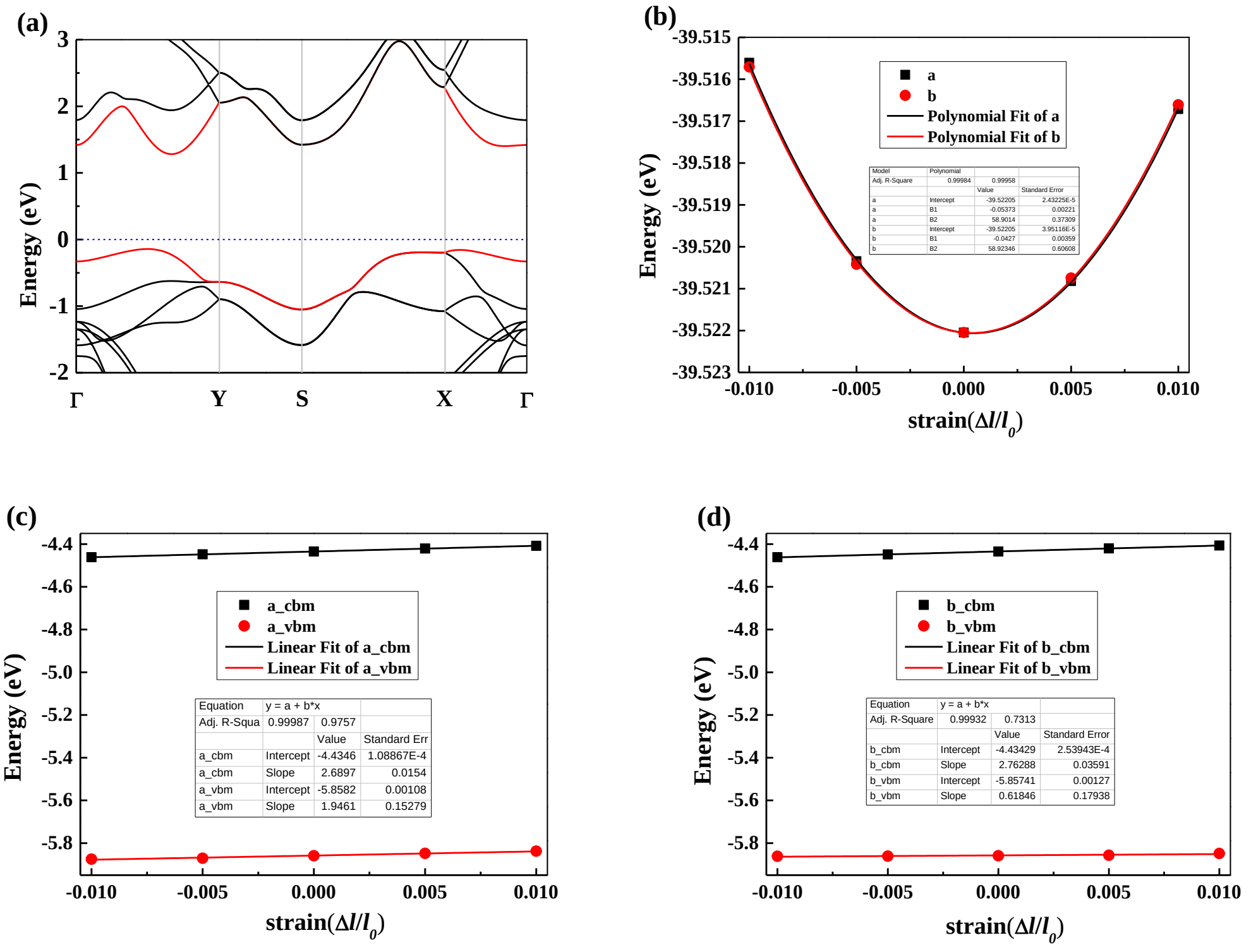

**Fig. S14** (a) Electronic band structure of SiS bilayer in an orthogonal supercell; (b) Total energy difference between unstrained and strained SiS bilayers along the *a* and *b* directions; (c) Energy shift of VBM and CBM for bilayer SiS with respect to the lattice dilation and compression along the *a* direction; (d) Energy shift of VBM and CBM for bilayer SiS with respect to the lattice dilation and compression along the *b* direction.

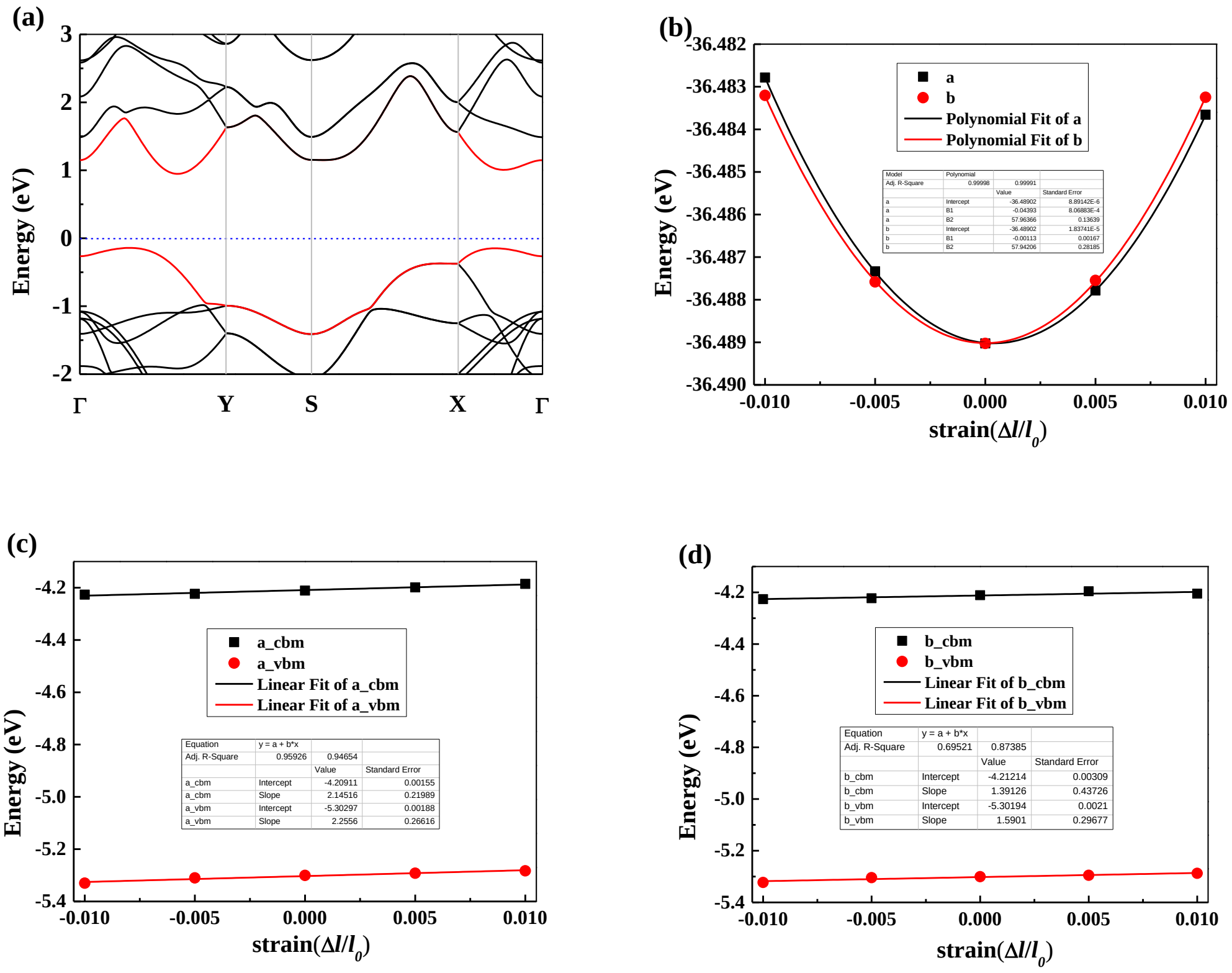

**Fig. S15** (a) Electronic band structure of SiSe bilayer in an orthogonal supercell; (b) Total energy difference between unstrained and strained SiSe bilayers along the *a* and *b* directions; (c) Energy shift of VBM and CBM for bilayer SiSe with respect to the lattice dilation and compression along the *a* direction; (d) Energy shift of VBM and CBM for bilayer SiSe with respect to the lattice dilation and compression along the *b* direction.